\documentclass[
 preprint,
 reprint, 
 superscriptaddress, 
 amsmath,amssymb,
 aps, prx, 
 floatfix,
]{revtex4-2}

\usepackage{graphicx}
\usepackage{xcolor}
\usepackage{dcolumn}
\usepackage{bm}
\usepackage{appendix}
\usepackage{upgreek}
\newcommand{\be}{\begin{equation}}
\newcommand{\e}{\end{equation}}
\newcommand{\beml}{\begin{subequations}}
\newcommand{\eml}{\end{subequations}}
\newcommand{\beq}{\begin{eqnarray}}
\newcommand{\eq}{\end{eqnarray}}
\newcommand{\ba}{\begin{array}}
\newcommand{\ea}{\end{array}}
\newcommand{\bpm}{\begin{pmatrix}}
\newcommand{\epm}{\end{pmatrix}}
\newcommand{\bc}{\begin{cases}}
\newcommand{\ec}{\end{cases}}

\begin{document}

\title{Magnetothermopower and particle-hole symmetry in a cuprate strange metal}

\author{Yu-Te Hsu}
\affiliation{High Field Magnet Laboratory (HFML-FELIX), Radboud University, Nijmegen, Netherlands}
\affiliation{Department of Physics and Department of Materials Science and Engineering, National Tsing Hua University, Hsinchu, Taiwan}
\affiliation{Center for Quantum Science and Technology, National Tsing Hua University, Hsinchu, Taiwan}

\author{Matija \v{C}ulo}
\affiliation{High Field Magnet Laboratory (HFML-FELIX), Radboud University, Nijmegen, Netherlands}
\affiliation{Institut za fiziku, Bijeni\v{c}ka 46, HR-10000, Zagreb, Croatia}

\author{Thom Ottenbros}
\affiliation{High Field Magnet Laboratory (HFML-FELIX), Radboud University, Nijmegen, Netherlands}
\affiliation{Institute for Molecules and Materials, Radboud University, Nijmegen, Netherlands}

\author{Yingkai Huang}
\affiliation{Van der Waals-Zeeman Institute, University of Amsterdam, Amsterdam, Netherlands}

\author{Jake Ayres}
\affiliation{High Field Magnet Laboratory (HFML-FELIX), Radboud University, Nijmegen, Netherlands}

\author{Steffen Wiedmann}
\affiliation{High Field Magnet Laboratory (HFML-FELIX), Radboud University, Nijmegen, Netherlands}
\affiliation{Institute for Molecules and Materials, Radboud University, Nijmegen, Netherlands}

\author{Mikhail I. Katsnelson}
\affiliation{Institute for Molecules and Materials, Radboud University, Nijmegen, Netherlands}
\affiliation{Wallenberg Initiative Materials Science for Sustainability, Uppsala University, 75121 Uppsala, Sweden}

\author{Nigel E. Hussey}
\affiliation{High Field Magnet Laboratory (HFML-FELIX), Radboud University, Nijmegen, Netherlands}
\affiliation{Institute for Molecules and Materials, Radboud University, Nijmegen, Netherlands}
\affiliation{H. H. Wills Physics Laboratory, University of Bristol, Bristol, United Kingdom}

\author{Mikhail Titov}
\affiliation{Institute for Molecules and Materials, Radboud University, Nijmegen, Netherlands}

\date{\today}

\begin{abstract}
Strange metallicity has been observed in several classes of quantum-critical metals and unconventional superconductors, including heavy-fermion compounds, iron-based superconductors, moiré materials, organic conductors, nickelates, and, most prominently, the high-$T_c$ cuprates. Although its principal properties – the ubiquitous linear-in-temperature resistivity, the hyperbolic magnetic-field scaling of the magnetoresistance and the anomalous reduction of the Hall carrier density – are widely regarded as signatures of non-Fermi-liquid transport, the microscopic nature of the charge carriers responsible for the strange-metal response remains unresolved. Here, we report magnetothermopower measurements on overdoped (Bi$_{2-x}$Pb$_x$)Sr$_{2-x}$La$_x$CuO$_{6+\delta}$ (Bi2201) single crystals in magnetic fields up to 35~T. Whereas the temperature dependence of the zero-field Seebeck coefficient $S(T)$ can be captured using Boltzmann transport theory, the field-dependent response $S(H)$ cannot. Instead, the magnetothermopower contains a large additional contribution whose field and temperature dependence is consistent \textcolor{black}{with the presence of short-range superconducting order well above $T_\textrm{c}$}. Combined with earlier Hall and magnetoresistance results, these data imply that the overdoped cuprate strange metal contains two transport sectors with distinct particle-hole symmetry: a conventional particle-hole-asymmetric Fermi-liquid (FL) contribution governing the Hall effect and the zero-field thermopower, and a nearly particle-hole-symmetric sector dominating the anomalous longitudinal magnetotransport. We formulate a phenomenological real-space model in which disconnected FL islands are embedded in a compensated Dirac liquid of phase-incoherent $d$-wave Bogoliubov quasiparticles. This picture reconciles conventional zero-field transport with anomalous magnetothermopower and magnetoresistance and offers a concrete framework for thinking about strange metallicity in overdoped cuprates.
\end{abstract}

\maketitle


\section*{Introduction}
Overdoped (OD) high-$T_\textrm{c}$ cuprates, i.\,e.~materials doped beyond the pseudogap endpoint $p^\ast$, were long regarded as the simplest part of the cuprate phase diagram: a regime in which coherent quasiparticles occupy a large Fermi surface and standard Fermi-liquid (FL) ideas should apply \cite{keimer2015}. In single-layer, hole-doped cuprates such as Bi2201, Tl$_2$Ba$_2$CuO$_{6+\delta}$ (Tl2201), and La$_{2-x}$Sr$_x$CuO$_4$ (LSCO), angle-resolved photoemission, quantum oscillations, and related probes reveal large Fermi surfaces and well-defined quasiparticles \cite{hussey2003, plate2005, yoshida2007, ding2019, smit2025, vignolle2008}, in broad agreement with first-principles band-structure calculations \cite{sakakibara2012}. This apparent simplicity, however, proved deceptive. Indeed, the normal state between $p^\ast$ and the superconductivity threshold $p_\textrm{sc}$ is now widely identified as a strange metal, characterized by a robust low-temperature $T$-linear resistivity that extends across the overdoped side of the phase diagram \cite{cooper2009, hussey2013, legros2019, putzke2021}. (Throughout, we distinguish this low-$T$ strange-metal regime from the high-$T$ bad-metal regime near room temperature \cite{hussey2004}.) These seeming contradictions have motivated an extensive effort to understand the superconducting (SC) and non-SC ground states of OD cuprates from both conventional and unconventional viewpoints \cite{pelc2019,zaanen2019,phillips2020,seibold2021,CuloDuffy-SciPost-2021,li2021,ayres2022,michon2023,patel2023,patel2024,liu2024,chang2024,bashan2026,ramshaw2025,fratini2026}.

A possible clue to the resolution of this problem is provided by the detailed decomposition of the longitudinal and transverse transport coefficients. Below $\sim 150$~K, the in-plane resistivity of OD cuprates can be well described by the expression $\rho_{ab}(T)=\rho_0+\alpha_1 T+\alpha_2 T^2$ \cite{cooper2009,hussey2013,harada2022}. In principle, such a form could arise from a single fluid with additive scattering rates, from two distinct fluids coupled effectively in series, or from some combination thereof. Recent high-field magnetotransport has sharpened this distinction. The in-plane magnetoresistance (MR) is found to follow a hyperbolic form, i.e.~$\rho_{ab}(T,H) = \mathcal{F}(T)+\sqrt{(\alpha_H T)^2+(\gamma_H H)^2}$, with $\mathcal{F}(T)=\rho_0+\alpha_T T+\alpha_2 T^2$ and $\alpha_1=\alpha_T+\alpha_H$ \cite{ayres2021, ayres2024}. This unusual MR scaling is incompatible with any straightforward extension of Boltzmann transport theory. By contrast, the Hall resistivity $\rho_{xy}(T,H)$ can be consistently accounted for within a Boltzmann treatment that assumes an anisotropic scattering rate $1/\tau_k$ \cite{putzke2021}, albeit with an effective Hall carrier density $n_{\rm H}$ that is below the total density expected from photoemission experiments\cite{putzke2021}. Taken together, these results imply that strange-metal transport contains one sector compatible with a quasiparticle description and another that does not \cite{ayres2022}.

Thermopower can, in principle, provide an incisive test of this dichotomy because, unlike the longitudinal resistivity, it directly probes particle-hole asymmetry. A finite zero-field Seebeck coefficient $S(T)$ and a finite Hall resistivity both require some degree of electron-hole imbalance, whereas a perfectly compensated sector contributes only weakly to either quantity at leading order. This observation motivates the present study of the magnetothermopower $S(T,H)$ in overdoped Bi2201. As with the Hall and MR responses, we find a pronounced duality: the zero-field thermopower is well captured by Boltzmann theory with an anisotropic scattering rate, whereas the field-dependent response is far larger than predicted within the same framework and evolves with field and temperature in a way that points to the predominance of superconducting (SC) fluctuations.

These data add a stringent constraint to the transport problem. If the bulk of the Hall response and the zero-field thermopower remain FL-like, while the longitudinal magnetotransport comes from another sector, then that second sector must be nearly particle-hole symmetric. A natural candidate is a phase-disordered $d$-wave paired state in which the pairing amplitude survives locally after long-range phase coherence is lost. In such a state, the low-energy nodal Bogoliubov quasiparticles remain Dirac-like and approximately particle-hole symmetric within the linearized nodal theory \cite{franzMillis1998,durstLee2000}. Of course, electron-hole symmetry of massless Dirac fermions plays a crucial role in electronic properties of graphene \cite{katsnelson_2020,castroNeto2009}. This symmetry structure immediately explains why the same sector can contribute strongly to the anomalous longitudinal field response while leaving a negligible imprint on $\rho_{xy}$ and on $S(H=0)$. The broader idea that pairing survives above the resistive transition is also consistent with spectroscopic evidence for pair formation above $T_{\rm c}$ in cuprates \cite{kondo2011,he2021,tromp2023}. 

Equally important is how the two sectors are combined. The transport phenomenology is not naturally organized by a simple momentum-space sum of parallel conductivities. Rather, it suggests a real-space inhomogeneous state in which a Hall-active, particle-hole-asymmetric FL component coexists with a nearly compensated sector that is much more visible in longitudinal field-dependent transport. In this paper we adopt the simplest nonpercolating-limit realization of that idea: disconnected FL islands embedded in a surrounding compensated Dirac liquid (CDL) of phase-incoherent Bogoliubov quasiparticles. The corresponding addition of longitudinal resistive contributions is used only as a qualitative limiting description; more generally, the problem is that of an inhomogeneous resistor network.
In the resulting transport topology, the FL islands dominate the Hall response and most of the zero-field thermopower, whereas the CDL dominates the anomalous magnetoresistance and magnetothermopower.

This interpretation is phenomenological by construction, but it is not microscopically implausible. Phase-fluctuation physics has long been argued to be important in cuprates with low superfluid stiffness \cite{emery1995}, and recent calculations for Hubbard-type models as well as spectroscopy near the overdoped edge of the SC dome point to a regime in which pairing survives while coherence becomes fragile \cite{kondo2011,he2021,tromp2023}. In particular, the $t-t'$ Hubbard model with parameters typical to 100-K SC cuprates show that the binding energy of two holes can be more than an order of magnitude higher than $T_c$, implying that incoherent Cooper pairs could indeed survive deep into the normal state \cite{Danilov2022,Stepanov2026}. This may indeed be sufficient to form Bogoliubov-like dispersion of quasiparticles, similar to the case of two-dimensional magnets where short-range magnetic order at finite temperatures results in an electronic structure similar to that found in fully ordered phases \cite{Irkhin1991}. Note that, in contrast to graphene, the Dirac point of these nodal Bogoliubov excitations is pinned to the Fermi level irrespective of the hole concentration or other parameters, which is crucial for realizing electron-hole symmetry in these systems.

A recent two-fluid, phase-separated model was also presented in the context of cuprates in Ref.~\cite{bashan2026}, though in that model, local SC droplets are dispersed within a FL host and act as scatterers of the normal carriers, rather than the other way around. We will show below that our model may account for the entirety of the transport data, not just the $T$-linear resistivity.  

The purpose of this paper is twofold. First, we show experimentally that the magnetothermopower of OD Bi2201 contains a large non-Boltzmann contribution that tracks SC fluctuations over a temperature window that is between 2 and 40 times the zero-field $T_c$ value. Second, we use these data to motivate a phenomenological two-fluid description of strange metallicity that links $S(T,H)$ to the previously established Hall and MR anomalies. This framework is intentionally minimal: it does not claim a complete microscopic theory of the strange metal. Rather, it provides a logically coherent way of organizing the transport phenomenology and of understanding how conventional and anomalous transport signatures can coexist within the same material.

\begin{figure*}[t]
\includegraphics[width=0.92\linewidth]{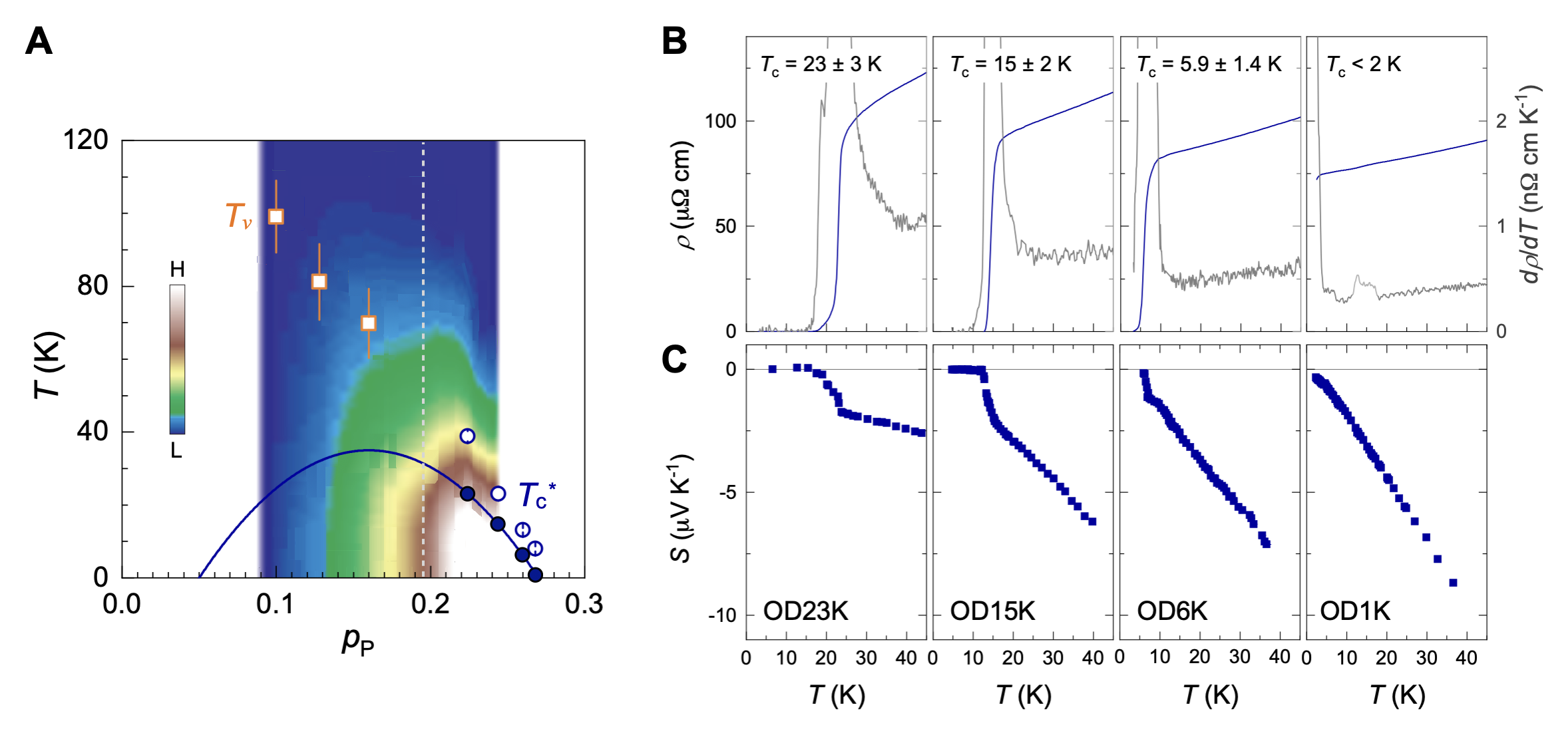}
\caption{Temperature-doping phase diagram of Bi2201 and transport characterization of the crystals studied in this work. (\textbf{A}) Characteristic temperatures of Bi2201 plotted against the hole doping level extracted using the Presland relation ($p_{\rm P}$). $T_{\nu}$ is the onset temperature of a finite Nernst signal observed in underdoped Bi2201 \cite{wang2006} and $T_{\rm c}^*$ is the onset temperature of SC fluctuations inferred from resistivity measurements. The color contour reflects the spectral weight associated with Cooper pairing previously determined using angle-resolved photoemission spectroscopy measurements \cite{kondo2011}. Vertical dashed line marks the location of pseudogap opening in Bi2201 \cite{berben2022}. The color contour and phase-boundary trends are adapted from Ref.~\cite{kondo2011} with permission. (\textbf{B}) Zero-field in-plane resistivity $\rho_{ab}(T)$ and (\textbf{C}) Seebeck coefficient $S(T)$ of the four Bi2201 crystals studied. $T_{\rm c}$ corresponds to the midpoint temperature of the resistive transition with the error reflecting the transition width. The four samples are referenced by their $T_{\rm c}$ values as OD23K, OD15K, OD6K, and OD1K.}
\label{zeroField}
\end{figure*}

\section*{Results}
\subsection*{Sample characterization}
Figure~\ref{zeroField} shows the zero-field in-plane resistivity $\rho_{ab}(T)$ and Seebeck coefficient $S(T)$ of four OD-Bi2201 crystals. A SC transition is found in all but the most OD crystal. $T_{\rm c}$ is defined as the midpoint of the resistive transition and is used for referencing all four samples (OD23K, OD15K, OD6K, and OD1K). SC transitions are also found in $S(T)$ for OD23K, OD15K, and OD6K. The $S(T)$ curves are all negative and decrease with increasing doping, consistent with previous reports \cite{kondo2005,lizaire2021}. For all samples, $S(T)$ varies linearly with $T$ within $T_{\rm c} \lesssim T \lesssim$~45~K, as expected for a metallic state at low $T$, though OD23K exhibits an increasing $|S/T|$ with decreasing $T$ \cite{lizaire2021}.

We define a temperature scale $T_{\rm c}^*$ for the onset of paraconductivity contributions to $\rho(T)$ as the temperature at which $d\rho/dT$ develops a clear upturn. The $T_{\rm c}^*$ and $T_{\rm c}$ values of the four crystals are plotted in Fig.~\ref{zeroField}A. $T_{\rm c}^*(p)$ is found to connect smoothly to the scale at which a sizable Nernst signal is found ($T_{\nu}$) in underdoped Bi2201 \cite{wang2006}. Note that here the SC dome of Bi2201 is assumed to follow the Presland relation \cite{presland1991} with a $T_c^{\rm max}$ = 36~K.

\subsection*{Zero-field $T$-dependent Seebeck coefficients}

\begin{figure*}[hbtp!!!]
\includegraphics[width=1\linewidth]{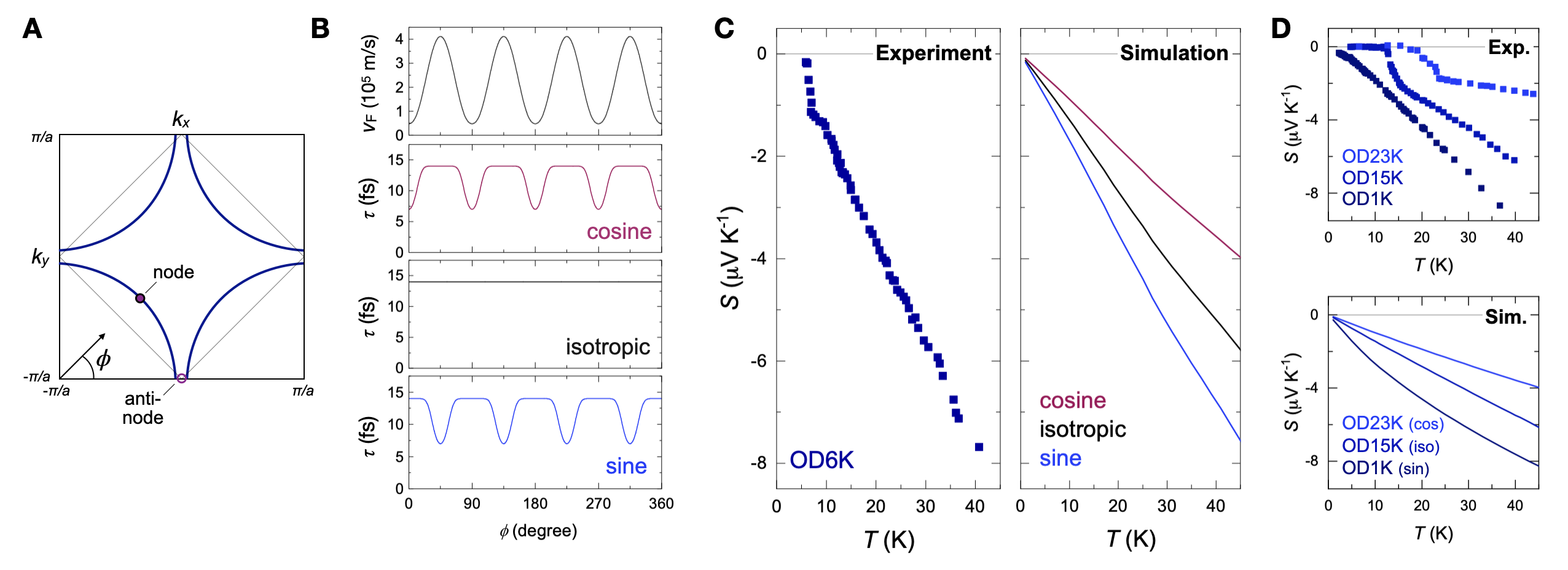}
\caption{Fermi surface parametrization and calculated Seebeck coefficients using the Boltzmann transport theory. (\textbf{A}) Fermi surface derived for OD6K using the tight-binding model described in \cite{kondo2005}. (\textbf{B}) Fermi velocity $v_{\rm F}$ and carrier lifetime  $\tau$ parametrizations as a function of azimuthal angle $\phi$. (\textbf{C}) $S(T)$ measured on OD6K (left) and simulated (right) based on the Fermi surface parametrization shown in (\textbf{A}), considering three cases of carrier lifetime anisotropy. (\textbf{D}) Experimental and simulated $S(T)$ of OD23K, OD15K, and OD1K. The form of $\tau$-anisotropy for each optimized simulation is specified in brackets. 
}
\label{boltzmannST}
\end{figure*}

The Seebeck coefficients are all negative, i.\,e. of opposite sign to the corresponding Hall coefficients \cite{putzke2021}. The negative sign of the Seebeck coefficients can be understood by considering the energy profile of the electrical conductivity, $\sigma(\epsilon)$, which has a maximum near the van Hove singularity (vHs) at $p_{\rm vHs} \approx 0.27$.
According to the Mott formula applicable to metals at low $T$ \cite{behnia_book}:
\begin{equation}
S_{\rm Mott} = \frac{\pi^2}{3}\frac{k_{\rm B}T}{e}\frac{\partial\ln\sigma_{\textit{xx}}(\epsilon)}{\partial\epsilon}\bigg|_{\epsilon=\mu} = \frac{\pi^2}{3}\frac{k_{\rm B}T}{e}\frac{\sigma_{xx}'(\epsilon)}{\sigma_{xx}}\bigg|_{\epsilon=\mu},
\end{equation}
therefore the sign of $S$ is determined by the energy-derivative of the electrical conductivity $\sigma_{xx}'(\epsilon)$. As the maximum of $\sigma(\epsilon)$ lies below the chemical potential for OD Bi2201, $\sigma_{xx}$ decreases as $\epsilon$ increases i.\,e. $\sigma_{xx}'(\epsilon) < 0$. As a result, the electronic structure-derived Seebeck coefficient for OD cuprates is generally negative.  Essentially, the Seebeck coefficient measures the particle-hole asymmetry of the low-energy excitation spectrum, as discussed recently in relevance to other OD cuprates \cite{jin2021,georges2021}. 

Previously, it was shown that $S(T)$ of OD Bi2201 can be largely reproduced by considering solely the electronic structure and assuming an isotropic carrier lifetime around the Fermi surface \cite{kondo2005}. Here, we consider several forms of carrier-lifetime anisotropy $\tau(\phi)$ and find that the normal-state $S(T)$ of OD6K below 40~K can be best reproduced using a combination of isotropic and anisotropic components, i.\,e.~$1/\tau = [1/\tau_0 + 1/\tau_k (\textrm{sin}(2\phi))^{12}]$ (Fig.~\ref{boltzmannST}C). We note that a similar form of scattering rate (with the same exponent) was recently deduced in Nd-doped La$_{2-x}$Sr$_x$CuO$_4$ (Nd-LSCO) at a doping level close to where the Fermi level crosses the vHs \cite{grissonnanche2021} and in the electron-doped cuprate La$_{2-x}$Ce$_x$CuO$_4$ (LCCO) beyond optimal doping \cite{duffy2025}. In Nd-LSCO, $1/\tau$ is maximal near the Brillouin zone boundary where the vHs is located \cite{grissonnanche2021} while in LCCO, $1/\tau$ peaks near the antiferromagnetic zone boundary \cite{duffy2025}. In Bi2201, our best simulation is obtained with $\tau$ maximized at the antinodes (zone boundary) and minimized at the nodes. This form of $\tau(\phi)$ is nevertheless qualitatively consistent with photoemission experiments on OD Bi2201 ($T_{\rm c}$ = 7~K) \cite{kondo2006}. OD6K is chosen for this comparison since it is used for the tight-binding model fitting in \cite{kondo2005} and should have the most accurate Fermi surface parameterization. While we find that the doping evolution of $S(T)$ can be qualitatively captured (Fig.~\ref{boltzmannST}D), the agreement between experimental and calculated values for other samples are not as good as for OD6K. The numerical agreement may be further improved by fine tuning the tight-binding parameters (including the shift in $\mu$) and/or the lifetime anisotropy parametrization (see Fig.~\ref{boltzmannST}D). In the following section, we will focus on OD6K for which the Boltzmann calculation yields the best experimental agreement.

\subsection*{$H$-dependence of Seebeck coefficients}

\begin{figure*}[hbtp!!!]
\includegraphics[width=0.8\linewidth]{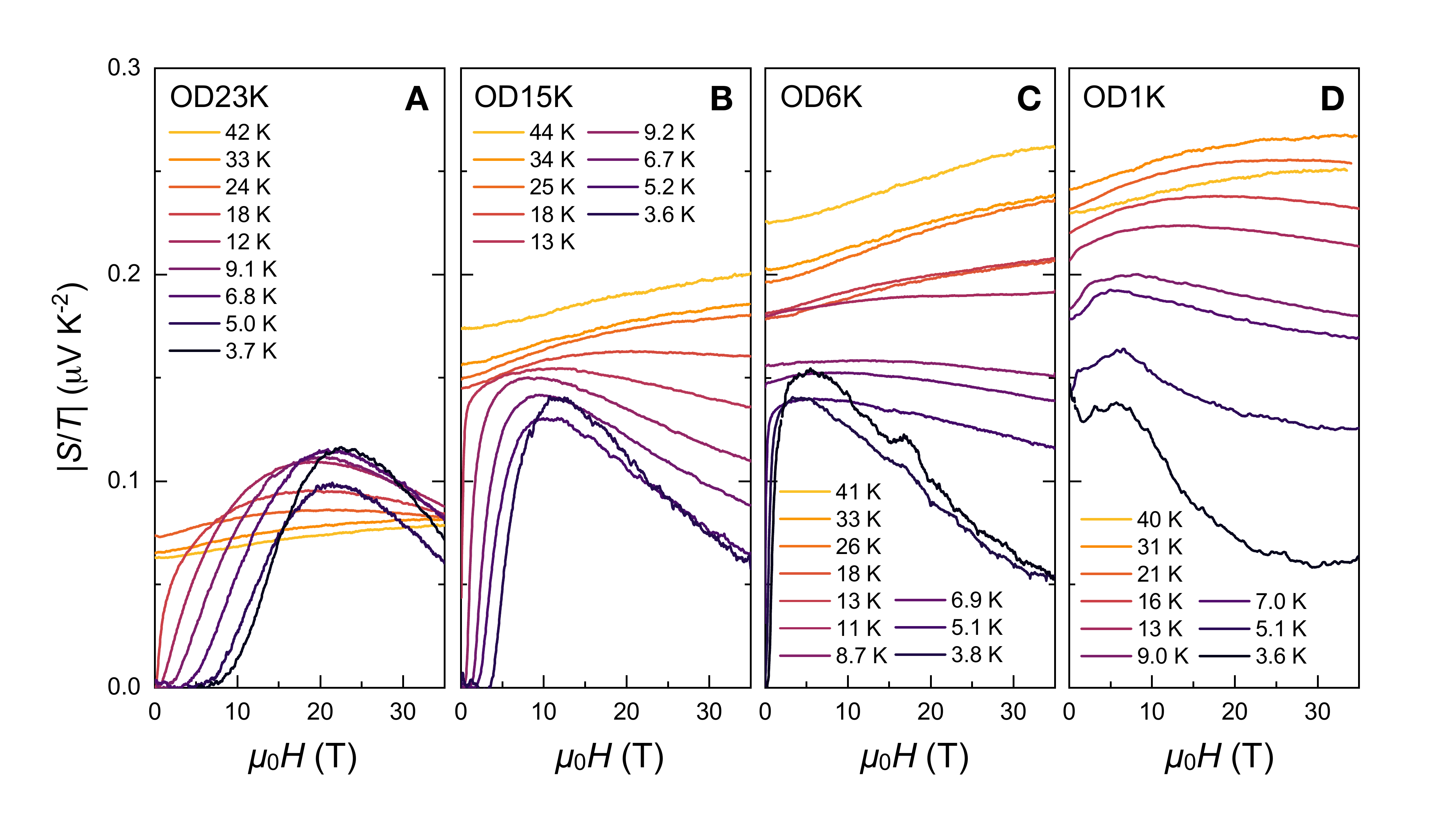}
\caption{Absolute Seebeck coefficient $|S/T|$ of OD Bi2201 measured in magnetic fields up to 35~T. Note that at high temperatures $|S/T|$ increases with magnetic field, whereas at low $T$, $|S/T|$ decreases with magnetic field in the high-field regime.}
\label{magnetoSeebeck}
\end{figure*}

Figure~\ref{magnetoSeebeck} shows the magnetic-field dependence of the absolute Seebeck coefficient $|S/T(H)|$ for all Bi2201 samples within the temperature range 3.6~K $\leq T \leq $ 45~K. A clear transition from a superconducting to a mixed state can be seen for all samples with a $T_{\rm c} > $~2~K. All samples exhibit a highly field-dependent $|S/T|$ with a peak structure that becomes more prominent with decreasing $T$. Notably, the field derivative of the magnetothermopower (MTEP) at 35~T is positive (i.\,e. $d|S/T|/d(\mu_0H) > 0$) at $T \sim$~40~K and becomes negative as $T$ decreases. Such a complex MTEP has been previously reported for heavy-fermion superconductor CeCoIn$_5$ \cite{onose2007} and interpreted as a rapid change in entropy flow in the precursor SC state. A similar peak structure in the MTEP has also been reported previously in LSCO \cite{collignon2021} and Bi2201 \cite{lizaire2021}.

\begin{figure*}[hbtp!!!]
\includegraphics[width=0.8\linewidth]{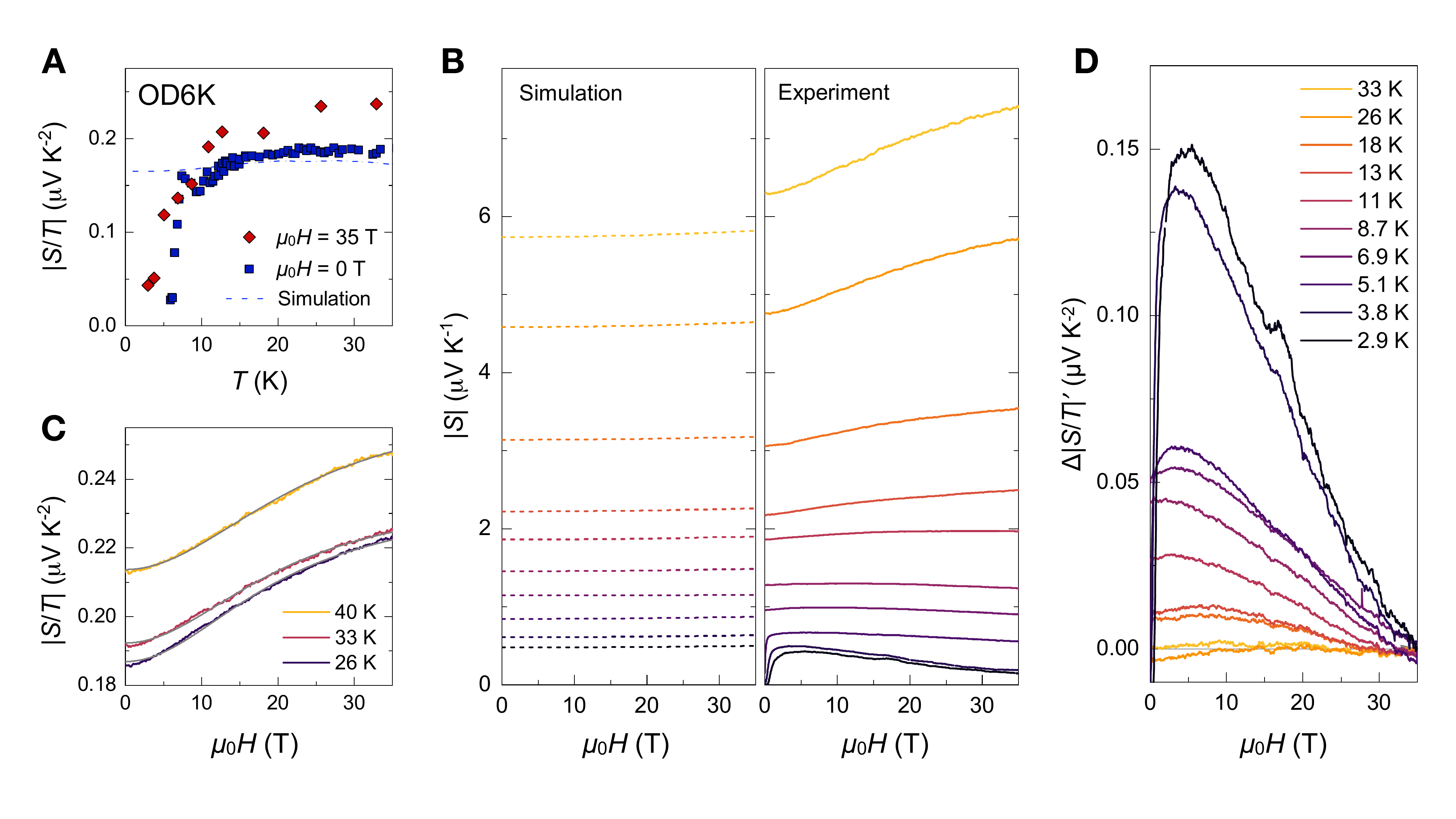}
\caption{Form of the normal-state magneto-Seebeck effect and discrepancy between electronic structure-derived and experimental $S(H)$ for OD6K.
(\textbf{A}) Absolute Seebeck coefficient $|S/T|$ measured in 0~T and 35~T. A good agreement between the experimental and electronic structure-derived $|S/T|$ is found at 0~T at $T > T_{\rm c}$, while $S/T$ measured in 35~T shows a sharp negative deviation from the zero-field behavior. (\textbf{B}) Left: Calculated $|S(H)|$ using the same set of parameters as in Fig.~\ref{boltzmannST}
Right: Experimental $|S(H)|$ which shows a pronounced and complex $H$ dependence with varying $T$. (\textbf{C}) Field-dependent $|S/T|$ at $T \gg T_{\rm c}$. All three traces show a similar dependence of $|S/T(H)|$ that can be described by the standard two-carrier form (shown in grey): $|S/T(H)|=AH^2/(1+BH^2)$. (\textbf{D}) The difference in $|S/T(H,T)|$ measured at specified temperatures and 40~K, $|\Delta S/T(T,H)| = |S/T(H,T)-S/T(H, 40~K)|$. For clarity, each $|\Delta S/T|$ trace is shifted so that its 35~T value equals zero.
}
\label{boltzmannSH}
\end{figure*}

In Fig.~\ref{boltzmannSH}A, we show $|S/T(T)|$ of OD6K measured at 0~T and 35~T. At 0~T, the measured $|S/T(T)|$ tracks closely to the simulated $|S/T(T)|$, showing a nearly $T$-independent behavior down to $T_{\rm c}^*$. The success of Boltzmann transport theory in describing $S(T)$ of OD Bi2201, however, contrasts markedly with our failure to capture the corresponding $S(H)$ within the same parameterization. As shown in Fig.~\ref{boltzmannSH}C, the simulated MTEP using the parameters deduced from fitting the zero-field thermopower is negligibly small compared to the experimental values. This diminished magnitude in the simulated $S(H)$ results from the large residual resistivity and a correspondingly short carrier mean free path $\ell \sim 3$~nm \cite{kondo2006}. Reproducing the magnitude of the experimental MTEP would require an artificial increase of $\ell$ by an order of magnitude, which will result in a residual resistivity that is unrealistically low. Furthermore, even an artificially enhanced $\ell$ will not yield a $T$-dependent sign change in the simulated MTEP. The experimental MTEP in OD Bi2201 thus motivates a description beyond Boltzmann transport theory and the relaxation time approximation.

Upon inspection, we find that the experimental MTEP measured at the three highest temperatures (26, 33, and 40~K) share a very similar form (Fig.~\ref{boltzmannSH}B). This form of MTEP is characteristic of a two-carrier longitudinal transport coefficient: $|S/T(H)=AH^2/(1+BH^2)|$, which can be well fitted to the experimental data. This observation suggests that the normal-state thermoelectric response in OD Bi2201 consists of two distinct terms. Indeed, the deviation from the high-$T$ normal-state MTEP behavior signifies the presence of an additional contribution to the Seebeck coefficient.

To examine the evolution of the MTEP with decreasing $T$, we show in Fig.~\ref{boltzmannSH}D the difference between the $|S/T|$ measured at each temperature and the 40-K value: $|\Delta S/T(T,H)| = |S/T(T,H)| - |S/T(40~\textrm{K}, H)|$. Note that for clarity all $|\Delta S/T(T,H)|$ traces are vertically shifted in such a way as to make its 35-T value equal to zero. The \lq residual' MTEP is found to be rapidly suppressed with increasing $H$, with its $T$ evolution bearing a strong resemblance to the Nernst coefficient $\nu$ measured in both underdoped and OD cuprates \cite{wang2006}, as well as in amorphous SC films of Nb$_{0.15}$Si$_{0.85}$ where the maximum in $\nu$ tracks the upper critical field below $T_c$ and the \lq ghost' critical field above it \cite{pourret2007}. In cuprates, a large Nernst response has long been associated with pseudogap physics and attributed to both SC phase fluctuations \cite{xu2000} and stripe correlations \cite{cyr-choiniere20009} inside the pseudogapped phase. In Bi2201, $p^*$ was recently determined in a combined transport and photoemission study carried out on the same batch of Bi2201 crystals \cite{berben2022}. As a result of that work, we can conclude that all samples investigated in this study lie beyond $p^*$ and that the sizable MTEP observed across the doping series can be attributed to SC fluctuations persisting far above $T_{\rm c}$ in OD Bi2201.

A strongly field-dependent MTEP was previously observed in electron-doped Pr$_{2-x}$Ce$_x$CuO$_4$ \cite{budhani2002} and qualitatively described by considering a flux-flow thermopower $S_{\rm f}$ comprising two terms (a quasiparticle component $S_{\rm qp}$ and a vortex component $S_{\rm vx}$ \cite{budhani2002,ri1993}:
\begin{equation}
S_\textrm{f}(H) =\rho_\textrm{f}S_\textrm{n}/\rho_n + \alpha s_\phi\rho_\textrm{f}/\phi_0 =\rho_\textrm{f}S_\textrm{n}/\rho_n + \alpha e_\textrm{N},
\label{Sf}
\end{equation}
where $\rho_{\rm f}$ is the flux-flow resistivity, $\rho_{\rm n}$ and $S_{\rm n}$ are respectively the resistivity and Seebeck coefficient in the normal state, $s_{\phi}$ is the transport entropy of the flux line, $\phi_0$ is the magnetic flux quantum, $\alpha$ is the vortex Hall angle, and $e_\textrm{N}$ is the Nernst coefficient. Here, $\rho_{\rm n}$ and $S_{\rm n}$ are assumed to be field-independent and $\alpha$ is assumed to be equal to the normal-state Hall angle. Since mobile vortices transport entropy along the thermal gradient, they also carry magnetic flux. This induces a (Nernst) voltage in the transverse direction that, when coupled via the Hall angle, can generate a finite longitudinal MTEP.

\begin{figure*}[hbtp!!!]
\includegraphics[width=0.8\linewidth]{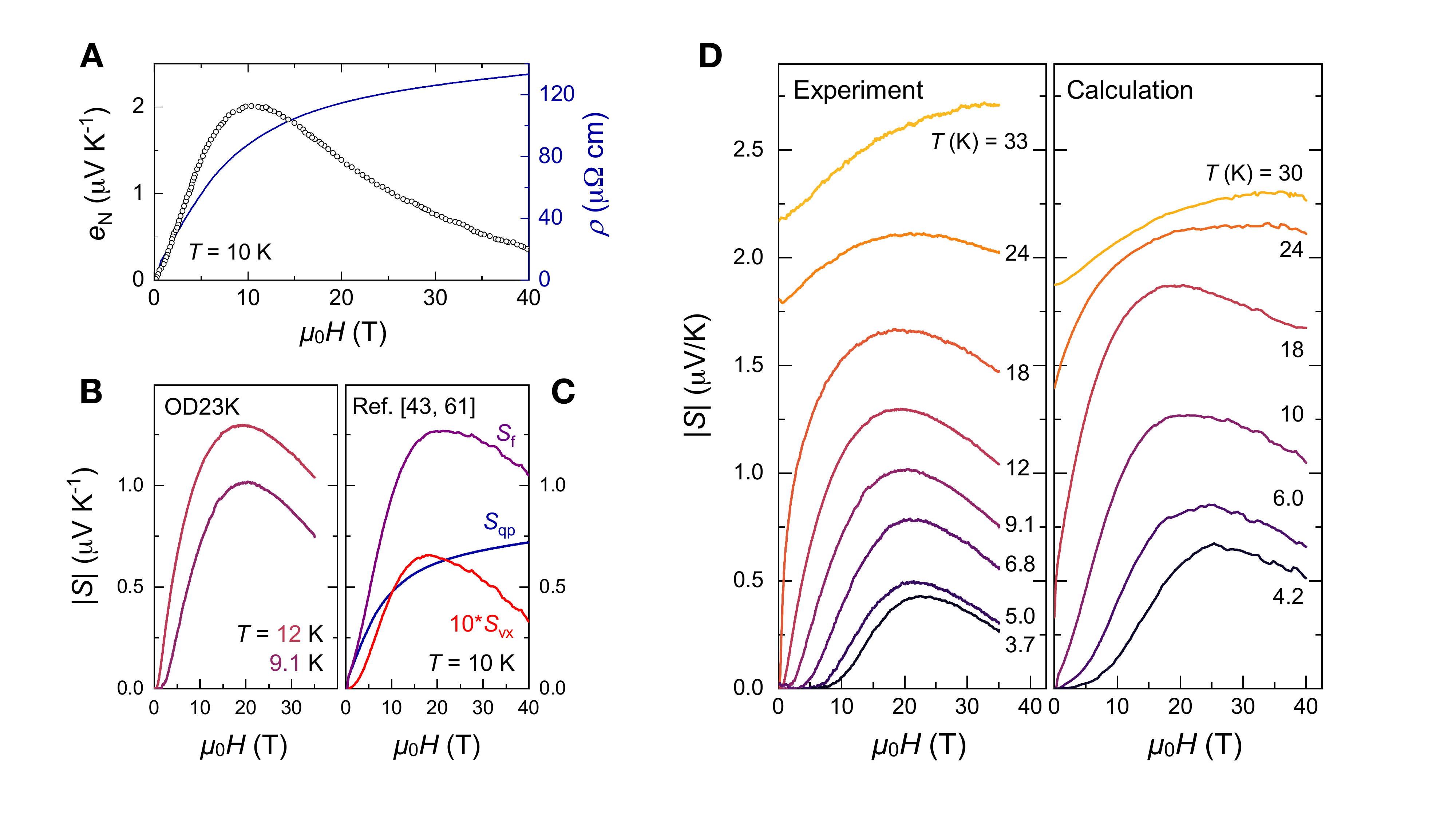}
\caption{Comparison between the measured magneto-Seebeck coefficient of OD23K and a calculation based on Eq.~(\ref{Sf}). (\textbf{A}) Nernst coefficient measured on an OD22K Bi2201 \cite{wang2006} and magnetoresistivity measured on an OD24K Bi2201 \cite{berben2022a} at $T$ = 10~K. (\textbf{B}) Absolute magneto-Seebeck coefficient $|S(H)|$ of OD23K at temperatures close to 10~K. (\textbf{C}) Calculated vortex-flow Seebeck coefficient $|S_{\rm f}(H)|$, together with its quasiparticle component $S_{\rm qp}$ and scaled vortex component $10 S_{\rm vx}$. The peak feature and negative slope at high magnetic fields in the measured $|S(H)|$ are well captured by the calculated $S_{\rm f}$. (\textbf{D}) Comparison of measured $|S(H,T)$ and calculated $|S_{\rm f}(H,T)|$ below 40~K using available data from \cite{wang2006,berben2022a}.
}
\label{vortexSeebeck}
\end{figure*}

In order to verify whether a vortex thermopower contribution can qualitatively describe the MTEP found in OD Bi2201, we use published Nernst ($T_{\rm c}$ = 22 K) \cite{wang2006} and MR data ($T_{\rm c}$ = 24 K) \cite{berben2022a} on OD Bi2201 crystals. As shown in Fig.~\ref{vortexSeebeck}A-C, $|S(H)|$ measured at $T \sim$ 10~K can be reproduced using Eq.~\ref{Sf}, provided a scaling factor of 8--10 is applied to $S_{\rm vx}$. This factor may arise due to the difference in the magnitude of the Hall angle $\alpha$ between the mixed and normal state. A similar scaling factor ($\alpha \sim 30$) was deduced previously in optimally doped YBa$_2$Cu$_3$O$_{6+x}$ \cite{ri1993}. Fig.~\ref{vortexSeebeck}D shows the same analysis applied to the common $(T,H)$ range in which the relevant data are available. Remarkably, despite the simplifying assumptions made in Eq.~\ref{Sf}, we find that the experimental MTEP can be qualitatively and semi-quantitatively reproduced using a thermoelectric response of a superconductor in the field-induced mixed state.

\section*{Summary of thermopower data}
Our study reveals that while the $T$ dependence of the zero-field Seebeck coefficient can be well described by the Boltzmann transport framework with known Fermi surface parameters, its $H$ dependence at $T<T_{\rm c}^*$ requires the incorporation of a vortex-like contribution to the MTEP. Given that the normal-state contribution in Eq.~\ref{Sf} can only increase monotonically with $H$, we conclude that the negative slope of $|S/T|$ must signify a finite contribution from short-lived Cooper pairs. While at present we cannot reliably extract the $H_{\rm c2}$ values due to the lack of Nernst data on these samples, the relatively high field strength required to fully suppress superconductivity in highly OD Bi2201 appears correlated with the constancy of the SC gap amplitude over the same doping region \cite{tromp2023}. Our current findings thus demonstrate that thermopower is a useful diagnostic tool for fluctuating superconductivity and reveals the unconventional nature of the breakdown of superconductivity in OD cuprates. 

Finally, the prominent SC fluctuation contribution to the MTEP implies that extracting the purely normal-state thermopower in OD cuprates requires considerable care. More importantly, it points to an OD regime in which the pairing amplitude remains robust even after long-range phase coherence has become fragile. Recent superfluid-density \cite{bozovic2016}, time-domain THz \cite{mahmood2019}, transport \cite{terzic2024}, disorder \cite{mahmood2022,juskus2024}, and scanning-tunneling studies \cite{li2022,tromp2023,ye2024} near the edge of the SC dome all point in the same direction. These developments motivate the two-liquid phenomenology introduced below, in which a phase-incoherent paired background coexists with a more conventional metallic sector.

\section*{Discussion}
The MTEP data sharpen the interpretation of the two-component magnetotransport phenomenology. In the same crystals, the zero-field Seebeck coefficient can be accounted for within a conventional Fermi-surface-based Boltzmann analysis, whereas the field-dependent thermopower and the previously reported quadrature MR require an additional sector that is largely invisible in the zero-field Seebeck and Hall responses. Within a single homogeneous fluid, this separation is difficult to rationalize. It is much more natural if the strange metal contains one sector that remains FL-like and particle-hole asymmetric, and a second sector that is nearly compensated and reveals itself primarily through the field-dependent longitudinal response.

\begin{figure}[tbh]
\includegraphics[width=0.85\columnwidth]{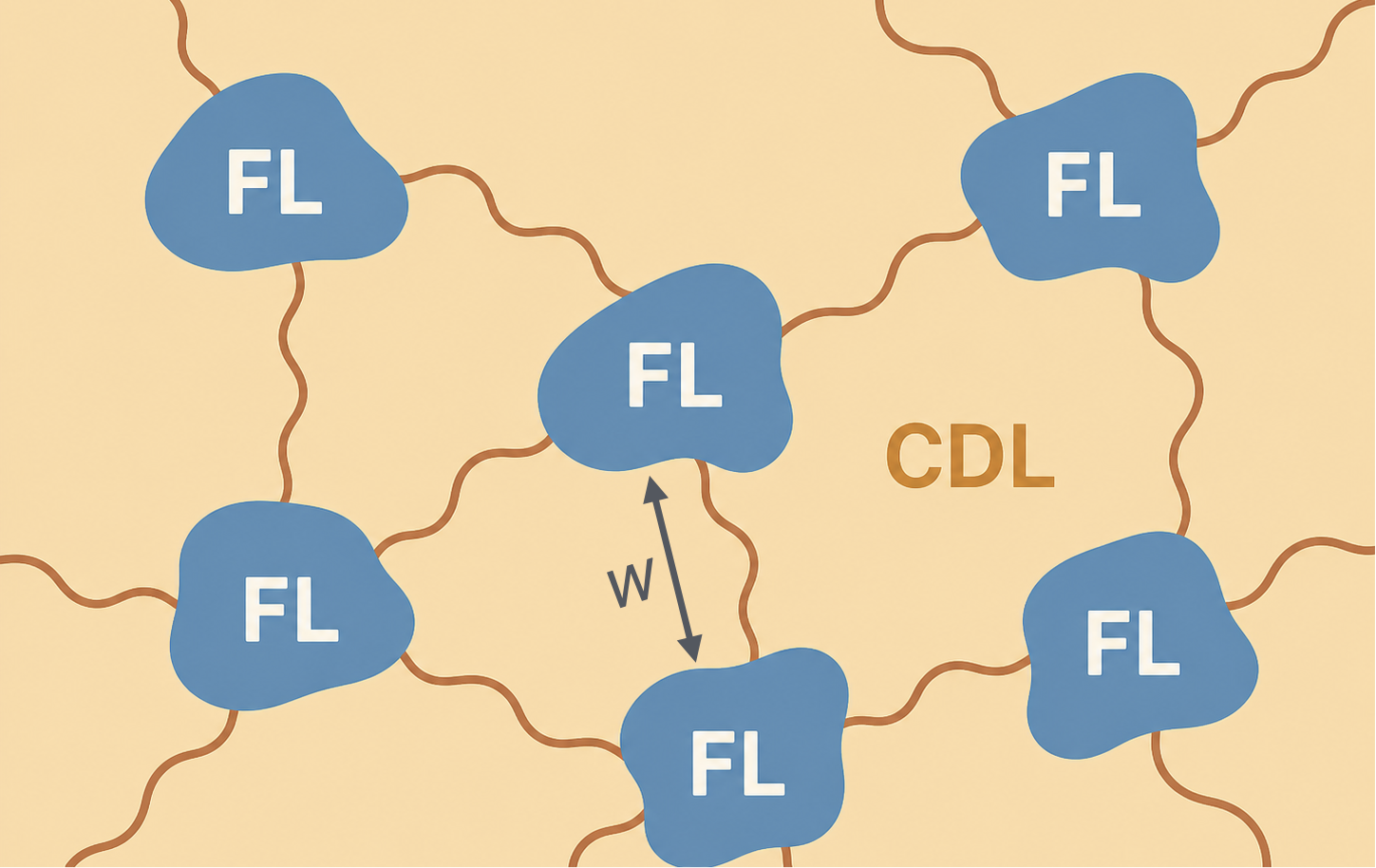}\\
\caption{Schematic real-space view of the proposed strange-metal regime. Wavy lines indicate fluctuating phase-slip trajectories that destroy long-range phase coherence for $T>T_\textrm{c}$. Disconnected Fermi-liquid (FL) islands are embedded in a phase-incoherent paired background that supports a compensated Dirac liquid (CDL) of nodal Bogoliubov quasiparticles. The characteristic spacing between neighboring FL islands is denoted by $W$. In the nonpercolating limit, charge transport over macroscopic distances necessarily involves current conversion through the CDL links between FL islands. The sketch is intended as a transport topology, not as a literal resistor-network solution.
}
\label{Fig_islands}
\end{figure}

\paragraph{Real-space composition and scope of the approximation.}
In a homogeneous multiband metal, distinct momentum-space sectors act as parallel channels and the conductivities add. The phenomenology inferred here is different: the Hall response and most of the zero-field thermopower remain FL-like, whereas the anomalous field-dependent longitudinal response requires an additional sector that contributes mainly to the longitudinal voltage drop. We therefore organize the strange-metal state in terms of a real-space texture in which disconnected FL islands are embedded in a surrounding compensated Dirac liquid (CDL), as sketched in Fig.~\ref{Fig_islands}. Note that this is very similar in spirit to the real-space picture deduced from a recent MR study on Bi2201 \cite{ayres2024}. 

In the nonpercolating limit, any macroscopic current path must repeatedly traverse the CDL links that separate neighboring FL islands. The total in-plane resistivity may then be written in the schematic form
\begin{equation}
\label{R}
\rho_{ab}(T,H)\simeq \rho_{ab}^{\rm FL}(T,H)+\rho_{ab}^{\rm CDL}(T,H).
\end{equation}

Eq.~(\ref{R}) is not intended as an exact composition law for an arbitrary inhomogeneous conductor, nor as a literal application of Matthiessen's rule. A real-space mixture of the type shown in Fig.~\ref{Fig_islands} is, strictly speaking, a random resistor network containing both series and parallel current paths. The additive-resistivity form in Eq.~(\ref{R}) should therefore be understood only as the simplest phenomenological limit appropriate when the FL regions are metallic but nonpercolating, so that the macroscopic longitudinal voltage drop must sample both sectors successively. A more quantitative treatment would require either numerical random-resistor-network modeling or an effective-medium construction of the Bruggeman type. For the purposes of the present paper, however, Eq.~(\ref{R}) is sufficient to encode the transport topology suggested by the data.

\paragraph{Physical content of the CDL.}
The CDL is not introduced as an independent metallic band. Rather, it is the phase-disordered fermionic sector of a $d$-wave paired state whose local pairing amplitude survives above the loss of macroscopic phase coherence \cite{emery1995,franzMillis1998,kondo2011,he2021,tromp2023}. In this sense, the CDL should be viewed as the fermionic descendant of superconductivity in a regime where Cooper pairs no longer carry dissipationless current, not as a conventional fluctuation-conductivity channel and not as an additional Fermi-surface sheet. The corresponding low-energy excitations are nodal Bogoliubov quasiparticles with spectrum
\begin{equation}
\label{X}
\epsilon_{\mathbf{k}}=\pm \sqrt{\xi_k^2+|\Delta_{\mathbf{k}}|^2},
\end{equation}
where the superconducting gap $\Delta_{\mathbf{k}}$ vanishes at the nodes. Near the nodes this spectrum is Dirac-like and, within the linearized nodal approximation, approximately particle-hole symmetric \cite{franzMillis1998,durstLee2000}. Formally, this makes the CDL analogous to transport near charge neutrality in a Dirac system such as graphene \cite{katsnelson_2020,castroNeto2009}, but with one crucial difference: here the Dirac sector is tied to a phase-incoherent $d$-wave background and coexists in real space with metallic FL regions.

This phenomenological picture is additionally motivated by recent microscopic work \cite{Danilov2022,Stepanov2026} indicating that pairing can survive far above $T_c$ while still losing coherence, in close analogy with the persistence of local magnetic moments above the ordering temperature in itinerant magnets \cite{moriya_book,Stepanov2022}. Above $T_\textrm{c}$, the pair amplitude is assumed to remain finite locally, while long-range phase rigidity is destroyed by fluctuations. The low-energy fermionic excitations of this background are then naturally described as nodal Bogoliubov quasiparticles with the spectrum of Eq.~(\ref{X}).

Within this picture, the FL islands correspond to regions where the local paired state is weakened strongly enough that the low-energy response reverts to a conventional metal. We do not attempt to identify a unique microscopic origin for these islands. They may be nucleated by strong out-of-plane impurity potentials, by local density fluctuations, or by the same mesoscopic fluctuations that destroy global phase coherence. What matters phenomenologically is that the FL component is metallic, comparatively particle-hole asymmetric, and spatially disconnected, whereas the surrounding CDL remains nearly compensated. This interpretation is particularly appealing for Bi2201, where disorder is substantial and mesoscopic inhomogeneity is already known to be important for the superconducting state \cite{tromp2023,ye2024}.

A brief clarification of terminology is useful here. Throughout this section, the term ``compensated'' refers to a \emph{strongly suppressed Hall field} of the CDL, i.\,e.~to the near cancellation of the electronlike and holelike Bogoliubov contributions in the linearized nodal theory. We do \emph{not} use the term in the narrower semimetallic sense of equal electron and hole Fermi-volume counts. In a generic two-band metal, equal densities alone do not guarantee a vanishing Hall signal; the mobilities also matter. In the present case, the small Hall response of the CDL is tied more fundamentally to the approximate particle-hole symmetry of the Bogoliubov spectrum. Beyond the strict nodal approximation, Fermi-surface curvature and electron-hole asymmetry can generate residual corrections, so the cancellation need not be exact; nevertheless, the leading Hall and zero-field Seebeck responses of the CDL are expected to be strongly suppressed.

\paragraph{Minimal drift-diffusion description of the CDL.}
To connect this phenomenology directly to transport, it is useful to write the simplest drift-diffusion equations for the CDL. Let $\mathbf{j}^{e}$ and $\mathbf{j}^{h}$ denote the electronlike and holelike quasiparticle currents, respectively. We define the charge-current combination $\mathbf{j}=\mathbf{j}^{e}-\mathbf{j}^{h}$, so that the electrical current density is $\mathbf{J}=-e\mathbf{j}$, and the neutral quasiparticle current $\mathbf{P}=\mathbf{j}^{e}+\mathbf{j}^{h}$, which in this simplified description is the current most directly associated with heat-carrying quasiparticles. We also decompose the quasiparticle density as $\rho=\rho_0+\delta\rho$, where $\rho_0$ is the equilibrium quasiparticle density of the CDL and $\delta\rho$ its nonequilibrium correction. This $\rho$ should not be confused with the sample resistivity. In terms of these quantities, the basic CDL equations can be written as
\beml
\label{Dirac}
\begin{align}
\label{Di1}
&\mathbf{P} =-\upmu \mathbf{j}\times\mathbf{B}- D \boldsymbol{\nabla}\delta\rho,\\
\label{Di2}
& \mathbf{j} = -\upmu \rho_0 \mathbf{E} -\upmu \mathbf{P}\times\mathbf{B},\\
\label{Di3}
&\mathrm{div}\,\mathbf{P}=-\delta\rho/\tau_\textrm{R},\qquad  \mathrm{div}\,\mathbf{j}=0.
\end{align}
\eml

Here $\upmu$ is the effective mobility of the CDL quasiparticles, $D$ is their diffusion coefficient, and $\tau_\textrm{R}$ is the recombination time of the Bogoliubov-Dirac quasiparticles. In the isotropic nodal approximation one has $\rho_0 \propto T^2/(\hbar v)^2$, where $v$ is a characteristic nodal velocity. If the dominant disorder acting on the phase-incoherent paired background is resonant scattering from phase slips or vortexlike defects, then $\tau \propto 1/|\epsilon|$ at low energies, implying $\upmu \propto 1/T^2$ when evaluated at thermal energies. The associated bare zero-field CDL resistivity is $r_0=(e\upmu\rho_0)^{-1}$.

Two remarks are important. First, Eqs.~(\ref{Dirac}) are introduced only as a minimal phenomenological model for the CDL sector; their purpose is to identify the structure of field-induced charge and neutral currents. Second, the experimentally fitted residual term of the \emph{whole sample} should not be identified with the bare CDL quantity $r_0$. The measured zero-field resistance generically contains contributions from the FL islands, the CDL links, and the FL/CDL conversion regions at their boundaries.

These equations already make the transport logic transparent. In zero magnetic field an ideal CDL contributes little to $\rho_{xy}$ and to $S(H=0)$ because its Hall and zero-field thermopower responses are suppressed by compensation, even though it can still contribute to the longitudinal resistance. In finite field, however, the Lorentz terms in Eqs.~(\ref{Di1}) and (\ref{Di2}) couple $\mathbf{j}$ and $\mathbf{P}$: a longitudinal charge current generates a transverse neutral quasiparticle flow, and a thermal drive generates a coupled electrical response. This is the basic mechanism through which the CDL can remain almost invisible in the Hall channel while becoming prominent in the longitudinal magnetotransport.

\paragraph{Implications for magnetoresistance and magnetothermopower.}
The same geometry that motivates Eq.~(\ref{R}) also provides a natural route to nonsaturating linear MR. Consider a CDL strip of width $W$ between neighboring FL islands. Because a neutral quasiparticle current cannot propagate through a conventional FL region without being converted, $\mathbf{P}$ must vanish at the FL/CDL boundaries and recombine over a characteristic length scale. Solving Eqs.~(\ref{Dirac}) in this geometry reproduces the recombination-limited linear-MR mechanism familiar from compensated two-component systems \cite{alekseev2015}. The relevant recombination length is
\begin{equation}
\ell_\textrm{R}=2\sqrt{\frac{D\tau_\textrm{R}}{1+(\upmu B)^2}},
\end{equation}
and in the field window
\begin{equation}
\label{L}
\ell_\textrm{R} \ll W\ll (\upmu B)^2\ell_\textrm{R}.
\end{equation}
The CDL contribution to the sheet resistance then becomes
\begin{equation}
\label{res}
R_\square= r_0 \frac{W}{\ell_\textrm{R}}
= \frac{r_0}{2}\frac{W\upmu |B|}{\sqrt{D\tau_\textrm{R}}}
= \frac{W |B|}{2 e \rho_0 \sqrt{D\tau_\textrm{R}}}.
\end{equation}
The purpose of Eq.~(\ref{res}) is qualitative: once current transport is bottlenecked by the conversion of a field-induced neutral flow at FL/CDL boundaries, a large linear MR follows naturally without invoking a conventional orbital mechanism.

The Hall and thermoelectric responses fit into the same logic. Because the linearized nodal CDL is nearly particle-hole symmetric, its intrinsic Hall conductivity and its zero-field Seebeck coefficient are parametrically small. The FL islands therefore dominate $\rho_{xy}$ and most of the zero-field thermopower, which explains why both quantities can still be described within an FL-like Boltzmann framework even when the MR and magnetothermopower cannot. In this interpretation, the smooth evolution of the low-$T$ Hall number across the strange-metal regime \cite{putzke2021} reflects the changing weight and connectivity of the FL sector as doping is increased toward $p_\textrm{sc}$.

The magnetothermopower is then not an isolated anomaly, but a thermoelectric signature of the same phase-incoherent sector. A thermal gradient drives a quasiparticle current $\mathbf{P}$ in the CDL; in finite field the Lorentz-coupling terms in Eqs.~(\ref{Di1}) and (\ref{Di2}) rotate part of this neutral flow into an electrical response. Because the CDL remains nearly compensated, the leading zero-field Seebeck contribution stays small, whereas the field-dependent longitudinal component can become large. This is consistent with the experimental observation that $S(T)$ at zero field looks broadly conventional while $S(H)$ becomes highly anomalous. It is also consistent with the close empirical resemblance between the magnetothermopower and vortex/Nernst phenomenology: the same phase slips that destroy superconducting coherence are natural candidates for the resonant scattering centers entering the CDL description. 

Finally, the present phenomenology also clarifies what can and cannot be inferred from doping trends. Increasing overdoping is expected to increase the fraction and/or connectivity of the FL sector, thereby shifting transport weight from the CDL-dominated longitudinal anomaly toward more conventional FL responses. It would be premature, however, to infer a unique doping dependence from the single parameter $W$: the Hall number, the zero-field thermopower, and the detailed shape of $S(H)$ depend separately on FL volume fraction, island morphology, boundary density, and recombination physics. A quantitative account of the doping evolution is therefore a task for future effective-medium or random-resistor-network modeling, not for the present discussion.

In summary, the two-liquid picture advanced here is intentionally phenomenological but internally consistent. Its purpose is not to provide an exact effective-medium theory of an inhomogeneous cuprate, but to explain why apparently conventional Hall and zero-field thermopower can coexist with a strongly anomalous longitudinal field response. Once the FL sector and the nearly compensated CDL are spatially separated, that coexistence is no longer paradoxical: the FL islands carry most of the Hall and particle-hole-asymmetric response, while the surrounding phase-incoherent CDL governs the anomalous MR and magnetothermopower.



The presented framework is deliberately phenomenological. In particular, the addition of longitudinal resistivities should be viewed as a nonpercolating-limit shorthand for an inhomogeneous transport network, not as an exact composition law. A more quantitative description will require random-resistor-network or Bruggeman-type effective-medium modeling, together with a more microscopic account of the FL-island morphology, the recombination scale, and the conversion of charge and neutral currents at FL/CDL boundaries. Even so, the present phenomenology organizes Hall, MR, Seebeck, and MTEP data within a single internally consistent transport logic and provides a concrete way of connecting those data to the increasingly strong evidence that pairing amplitude and phase fluctuations survive deep into the overdoped regime \cite{ayres2021,ayres2024,putzke2021,rourke2011,he2021,tromp2023, terzic2024}. While our MTEP study is restricted to $T <$ 50 K, we note that the anomalous MR scaling -- that forms a critical part of this collective phenomenology -- can extend up to 200~K \cite{berben2022a}.

The picture advanced here therefore suggests several direct tests. Zero-field Hall transport and thermopower should track the growing weight and connectivity of the FL sector, whereas the MR and high-field magnetothermopower should correlate more directly with independent measures of phase fluctuations and spatial inhomogeneity. Comparative Hall, Nernst, and magnetothermopower measurements on the same crystals, combined with local probes of disorder and gap inhomogeneity, should provide especially stringent tests of this emergent two-liquid description of overdoped cuprate strange metallicity.

\section*{Materials and Methods}
Bi2201 single crystals were grown using the floating-zone method and cleaved into rectangular pieces with typical dimensions of ($1500\times400\times10$)~$\mu$m$^3$. 25 or 50 $\mu m$-diameter gold wires were then attached to the samples using DuPont 6838 silver paint and heated at 425~$^{\circ}$C for 20 minutes to achieve contact resistance of $\sim$1~$\Omega$.

The in-plane electrical resistivity was measured in a standard four-point configuration using an ac excitation current with a frequency of 13--30~Hz and an amplitude of 0.5--1~mA. The Seebeck coefficient was measured using a standard dc two-thermometer-one-heater method with a home-built apparatus as described previously \cite{arsenijevic2016}. Ruthenium oxide chips with a room-temperature resistance $R \approx$~3.3~k$\Omega$ were used as both the heater and thermometer, calibrated \textit{in situ} by measuring the individual $R(T)$ curves between 2 and 60~K for each run. The sample temperature was controlled using a continuous He-flow cryostat under high vacuum. Thermoelectric voltages are measured using 25-$\mu$m phosphor bronze wires and a N11a nanovoltmeter by EM Electronics. Magnetic fields up to 35~T were applied along the  crystallographic $c$-axis using a Bitter magnet at the HFML-FELIX institute at Radboud University. Measurements were performed at selected temperatures in both field polarities and symmetrized to remove any possible contamination of the longitudinal magnetothermopower from the transverse Nernst signal (see Fig.~7).

For numerical calculations of the transport coefficients, we adopted the tight-binding model developed for Bi2201 as described in \cite{kondo2005}. Seebeck coefficient calculations were done using the Boltzmann formalism:
\begin{equation}
S = \frac{\alpha_{xx}}{\sigma_{xx}} = \frac{1}{|e|T}\frac{\int^\infty_{-\infty}d\epsilon\,
\sigma_{xx}(\epsilon)(\epsilon-\mu)\left(\partial f / \partial\epsilon\right)}
{\int^\infty_{-\infty}d\epsilon\,\sigma_{xx}(\epsilon)\left(\partial f/\partial\epsilon\right)},
\end{equation}
where $\sigma_{xx}$ is the longitudinal electrical conductivity, $\alpha_{xx}$ is the longitudinal thermoelectric conductivity, $\epsilon$ is the particle energy, $\mu$ is the chemical potential, and $f$ is the Fermi-Dirac distribution. The window in $\epsilon$ for the numerical integration is chosen to be $\mu \pm 10k_{\rm B}T$.
$\sigma_{xx}$ is calculated using the Jones-Zener expansion up to the second order and integrated over the entire Fermi surface:
\begin{equation}
\sigma_{xx}^{(n)} =- e^2\! \int\! \frac{\textrm{d}^3\textrm{k}}{4\pi^3\hbar}\, 
v_x\!\left[-\tau[\mathbf{v_k}\times\mathbf{B}]\cdot\frac{\partial}{\partial\mathbf{k}}\right]^n\!\! v_x \tau\, \frac{\partial f}{\partial\epsilon},
\end{equation}
where $v_x$ is the $x$-component of the Fermi velocity, $\tau$ is the quasiparticle lifetime, and $B$ is the magnetic field.
We parameterized the anisotropic carrier lifetime using:
\begin{equation}
    \tau = [1/\tau_0 + (1/\tau_k)F_{\tau_k}(2\phi)]^{-1},
\end{equation}
where $\phi$ is the angle within the $k_x-k_y$ plane, $1/\tau_0$ is the isotropic scattering rate, $1/\tau_k$ defines anisotropy strength, and $F_{\tau_k}$ is the function defining the form of scattering rate anisotropy. An equal value of $\tau_0 = \tau_k$ = 14~fs is used as input \cite{kondo2006}, yielding a residual resistivity $\sim 100~\mu\Omega$~cm consistent with the experimental values.

\section*{Acknowledgements}
We thank S. Badoux, K. Behnia, A. Georges, G. Grissonnanche, and S. Kasahara for fruitful discussions.


\section*{Supporting Information}
\begin{figure}[hbtp!!!]
\includegraphics[width=1\linewidth]{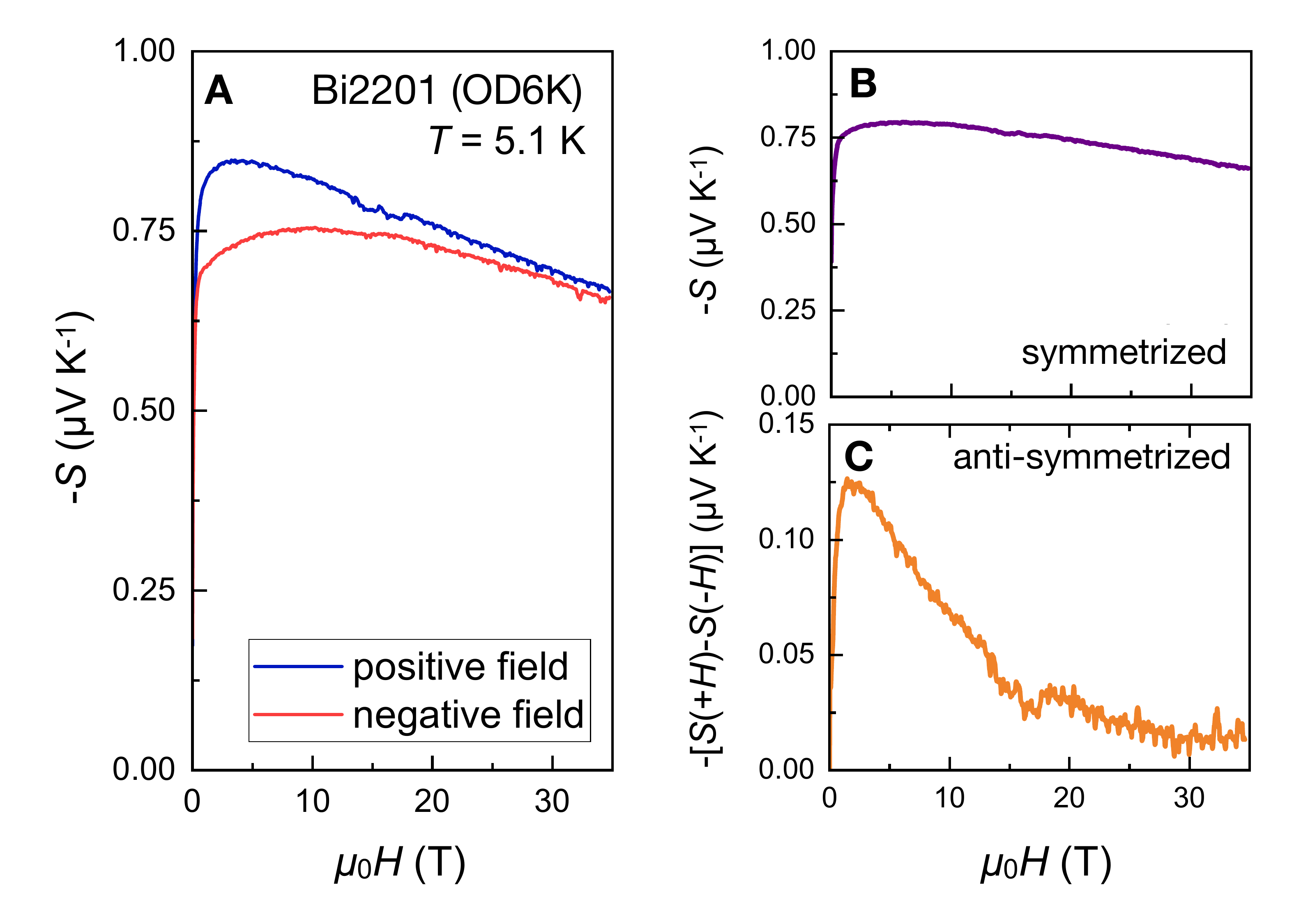}
\caption{Symmetrization procedure for the magnetothermopower data. (\textbf{A}) Unsymmetrized traces measured on OD6K at $T=5.1$~K for opposite field polarities. (\textbf{B}) Symmetrized magneto-Seebeck coefficient obtained by averaging the two traces in (\textbf{A}). (\textbf{C}) Antisymmetric component obtained from half their difference. This residual signal is attributed primarily to admixture of the transverse thermoelectric response due to contact misalignment and is expected to scale with the Nernst signal.}
\label{symmetrization}
\end{figure}

To isolate the longitudinal magnetothermopower, all field sweeps were symmetrized with respect to field reversal. Figure~\ref{symmetrization} illustrates the data-processing procedure. Each measurement was performed using the sequence 0 $\rightarrow$ +35~T $\rightarrow$ 0 $\rightarrow -35~$T $\rightarrow$ 0. We refer to the first half-cycle as the ``positive-field'' trace and to the second as the ``negative-field'' trace. Any difference between the two traces arises predominantly from a small admixture of the transverse thermoelectric voltage due to unavoidable contact misalignment. Because this admixture is odd in magnetic field, it is removed by averaging the positive- and negative-field traces. All magnetothermopower data shown in this work were processed in this way.

\makeatletter
\def\bibsection{\section*{References}}
\makeatother

\bibliography{reference}

\end{document}